# Epitaxial Growth of Quasi-One-Dimensional Bismuth-Halide Chains with Topological Non-Trivial Edge States

Jincheng Zhuang[1], Jin Li[2], Yundan Liu[2], Dan Mu[2], Ming Yang[1], Yani Liu[1], Wei Zhou[3], Weichang Hao[1], Jianxin Zhong[2], Yi Du[1]

[1] School of Physics, Beihang University, Haidian District, Beijing 100191, China

[2] Hunan Key Laboratory of Micro-Nano Energy Materials and Devices, and school of physics and optoelectronics, Xiangtan University, Hunan 411105, PR China

[3] School of Electronic and Information Engineering, Changshu Institute of Technology, Changshu, 215500, China

Jincheng Zhuang, Jin Li, and Yundan Liu contributed equally to this work. Correspondence and requests for materials should be addressed to J.C.Z. (email: jincheng@buaa.edu.cn) or to J.X.Z (jxzhong@xtu.edu.cn) or to Y.D. (email: ydu@uow.edu.au)

Quantum spin Hall insulators have one-dimensional (1D) spin-momentum locked topological edge states (ES) inside the bulk band gap, which can serve as dissipationless channels for the practical applications in low consumption electronics and high performance spintronics. However, the clean and atomically sharp ES serving as ideal 1D conducting channels are still lack. Here, we report the formation of the quasi-1D $Bi_4I_4$ nanoribbons on the surface of Bi(111) with the support of the graphene-terminated 6H-SiC(0001) and the direct observations of the topological ES at the step edge by scanning tunneling microscopy and spectroscopic-imaging results. The ES reside surround the edge of $Bi_4I_4$ nanoribbons and exhibits remarkable robustness against non time reversal symmetry perturbations. The theoretical simulations verify the topological non-trivial character of 1D ES, which is retained after considering the presence of the underlying Bi(111). Our study supports the existence of topological ES in $Bi_4I_4$ nanoribbons, paving the way to engineer the novel topological features by using the nanoribbons as the 1D building block.

As the typical representatives of two-dimensional (2D) topological insulators, quantum spin Hall (QSH) insulators feature conducting edge states (ES) with two spin-momentum locked channels under opposite directions, which are topologically protected from backscattering by time reversal symmetry (TRS)[1-4]. The fault-tolerance ability and concurrent two spin states makes ES of QSH insulators promising in future applications of low dissipation electronics, high performance spintronics, quantum computation, etc. Although there are plenty of candidates for the QSH insulators predicted according to the Kane-Mele theory or Bernevig-Hughes-Zhang theory[2,5], very few have been confirmed experimentally to be topological non-trivial. Until now, the evidence could only be identified in HgTe/CdTe and InAs/GaSb quantum wells[6,7], elemental Bi system[8-10], stanene[11,12], monolayer transition metal dichalcogenide materials in the distorted octahedral 1$T$' phase[13-17], and Dirac semi-metal thin film[18,19]. Nevertheless, the one-dimensional (1D) ES and the bulk states are bridged by the covalent or ionic bonds, leading to the shape and size of the 1D ES be confined by these of the 2D QSH insulators. Consequently, the fabrications of clean and atomically sharp edges, such as nanoribbons, which are essential for the observation of topological ES as well as serving as ideal 1D dissipationless conducting wires, are challenging.

The bismuth halide, $Bi_4X_4$ ($X$ = I or Br), has one-dimensional infinite molecular chain as its building block, where their three-dimensional (3D) materials are stacked by 1D chains through van der Waals force along both of the in-plane and out-of-plane directions. Furthermore, abundant topological states have been theoretically predicted and subsequently experimentally supported in this unique system. The monolayer $Bi_4Br_4$ was foretold to be a large-gap QSH insulator[20], and 1D helical hinge states of high-order topological insulators, resulted from the mutual effects of $C_2$ rotation symmetry and TRS, have been detected in its $α$' bilayer phase by laser angle-resolved photoemission spectroscopy (ARPES)[21]. For $Bi_4I_4$, the laser-ARPES results show that a crystal transition from the $β$-phase to the $α$-phase drives a topological phase transition from a weak topological insulator (WTI) (regarded as 3D allotrope of QSH insulators) to a normal insulator even at room temperature[22,23]. The first-principle calculations imply that the stain could further tune the topological properties of $β$-phase $Bi_4I_4$ ranging from WTI to a strong topological insulator and then to a Weyl semimetal and a normal insulator[24]. All these reports manifest that monolayer $Bi_4X_4$ system is a QSH insulator, and the exotic topological phases could be realized by introducing different kinds of symmetry-breaking perturbations. Therefore, $Bi_4X_4$ system is an appropriate platform to engender quasi-1D topological conducting wires with clean and atomically sharp edges. The conventional mechanical exfoliation methods, however, is dedicated to dimensionality reduction from 3D to 2D rather than 2D to 1D, leading to the formation of quasi-1D $Bi_4X_4$ still be unexplored.

In this work, we report the successful realization of quasi-1D $Bi_4I_4$ nanoribbons on Bi(111) surface with the support of graphene-terminated 6H-SiC(0001) substrate by molecular beam epitaxy method. The superstructure of $Bi_4I_4$ nanoribbons evoked by the substrate-nanoribbon interaction is identified by using the scanning tunneling microscopy (STM). The ES with the decay length around 1 lattice periodicity have been observed experimentally according to scanning tunneling spectroscopy (STS) technique. The density functional theory (DFT) calculations reveal that the topological invariant in the superstructure is retained compared to that of free-standing $Bi_4I_4$, confirming the non-trivial topology of the ES. Our work provides an effective way to synthesize the quasi-1D topological nanoribbons, benefiting not only to the fundamental research for the realization of exotic quantum states but also to the potential applications in nanoelectronics and spintronic devices.

The bismuth thin film displays the thickness dependent facet on the surface of graphene at room temperature, where a crystalline transition from (110)-orientated facet with the black phosphorus-like structure to (111)-orientated facet with the buckled honeycomb structure occurs after the layer number increases up to a critical value[25]. The similar growth dynamics have also been identified in our results, as shown in Supplementary Fig. 1. After the formation of the large-scale Bi(111) thin film, the element I is imported combined with the Bi during the following deposition process. The nanosized stripes are formed and arranged on the surface of Bi(111), as displayed in Fig. 1b. The high resolution STM image of the parallel stripes shown in Fig. 1d reveals a slightly zigzag arrangement of protrusions along the direction of the stripe. The distance between the nearest-neighboring protrusions is 4.2 Å, which is smaller than the periodicity (~ 4.4 Å) of free-standing $Bi_4I_4$ along the chain direction (vector ***b***). There is a displacement for the nearest-neighboring stripes along the direction perpendicular to the stripe, leading to every other stripe recovers its position. Thus, the unit cell of the stripe phase is labelled by the red rectangular frame in Fig. 1d with the lattice constant of 14.3 Å and 8.2 Å cross the stripes and along the stripe direction, respectively. It should be noted that the value of 14.3 Å is almost the same to the *a*-axis lattice constant of free-standing $Bi_4I_4$, while 8.2 Å is less than the two times the periodicity along *b*-axis. In order to resolve the detailed atomic structure of the striped phase, we made a comparative study of structural properties between the striped phase and the underlying Bi(111) substrate from the view of both periodicity and lattice orientation. The periodicity of Bi(111) thin film is around 4.7 Å, as shown in Fig. 1c, which is larger than the lattice constant (~ 4.54 Å ) of

7-layer Bi(111) thin film grown on graphene/SiC substrate but comparable to that (~ 4.75 Å ) of monolayer Bi(111) on single-layer Bi(110) with the support of highly oriented pyrolytic graphite[25,26]. Thus, the various periodicities of Bi(111) surface are attributed to the interfacial strain effect induced by the different sample thickness in these reports. Note that the lattice constants of the striped phase, 14.3 Å and 8.2 Å for the two lattice orientations, fits well with 3 times (3 × 4.7 Å = 14.1 Å) and √3 times (√3 × 4.7 Å = 8.2 Å) the periodicity of Bi(111), respectively. Furthermore, the directions cross the stripes and along the stripe are correlated to [$\bar{1}$10] direction (lattice orientation) and [$\bar{1}\bar{1}$2] direction (perpendicular to the lattice orientation) of the Bi(111) surface, respectively. Based on the experimental results, the structural model is thus proposed as follows: the $Bi_4I_4$ nanoribbon is adsorbed on the Bi(111) surface, where the lattice constant for the 3 ×√3 unit cell of Bi(111) matches that of the 1 × 2 superlattice in terms of free-standing monolayer $Bi_4I_4$. To build the coperiodic lattice of the $Bi_4I_4$/Bi hybrid system, $Bi_4I_4$ is compressed in the *b* direction by 7.4%, leading to the slightly zigzag-like stripe in our sample instead of the long straight chains in free-standing $Bi_4I_4$. Detailed information on structural properties and simulated STM images of both free-standing $Bi_4I_4$ and $Bi_4I_4$ on Bi(111) surface is given in Supplementary Fig. 2. The evolution from the atomic structure model to the experimental STM image of $Bi_4I_4$ on Bi(111) and then to the experimental STM image of Bi(111) in Fig. 1e schematically illustrate the structural relationship between $B_4I_4$ stripes and Bi(111), and indicates the agreement between the structural model and the experiment. The commensurate superstructure could be further confirmed by the stripe number evolution, as displayed in Fig. 1f and Supplementary Fig. 3, where the number of the stripes increases or decreases with the step value of two. In the free-standing $B_4I_4$, every chain is equivalent to each other as the in-plane interaction strength is uniform, resulting in the width value of the chains be randomly distributed. Since there are two chains of $Bi_4I_4$ in one 3 × √3 superstructure of Bi(111), the Bi(111)-$Bi_4I_4$ interface interaction gives rise to the simultaneous growth of two chains in addition to the existed stripes more energy favorite. Fig. 1a schematically illustrates the sandwich-ilke structure of our system with a $Bi_4I_4$ nanoribbon on top of Bi(111)/graphene/SiC(0001).

It has been reported that the 3D *β*-$Bi_4I_4$ is the WTI with a quasi-one-dimensional Dirac topological surface state at the side surface (perpendicular to (001) plane) identified by laser-ARPES[22]. Since the WTI could be constructed by stacking the weakly coupled layers of 2D topological insulators, the monolayer $Bi_4I_4$ is expected to be the QSH insulator with 1D topological ES in contract to the bulk insulating bulk. To unravel its electronic structure, we acquire tunneling spectra, which is proportional to the local density of states (LDOS), taken at the points locating exactly at edges (labelled by black point 1) and at inner terrace (labelled by green point 2) of the $Bi_4I_4$ nanoribbon shown in Fig. 2a, respectively. The black curve in Fig. 2b represents the bulk LDOS of $Bi_4I_4$, which features a gap-like dispersion with the bias voltage ranging from -200 mV to 160 mV. It is noticeable that the laser-ARPES results reveal that the valence band maximum (VBM) of 3D $\beta$-$Bi_4I_4$ is also around -200 mV[22], consistent with our STS spectra. The small residual tunneling density of states within the gap can be attributed to tunneling/coupling to the underlying Bi(111), which has been evidenced in the similar phenomenon of nonzero density of states in the gap in other QSH insulators, such as single-layer $WTe_2$ and $WSe_2$ on graphene[14,16]. In contrast to the gap in the bulk, d$I$/d$V$ spectrum measured at the edge of the nanoribbon (blue curve in Fig. 2b) displays the distinct 'V-shape' dispersion with states filling in the bulk gap, implying the existence of a conductive ES[10,14,27]. The real space spectroscopic mapping of a $Bi_4I_4$ nanoribbon with two straight edges along $b$-axis direction at the energy window of the bulk gap was performed to clarify the dispersion and the nature of the ES. The ES is localized and run continuously along our sample edges with a spatial extension close to the periodicity of $a$-axis (~1.5 nm), as labelled by the white dashed lines in Fig. 2c, which is comparable to the penetration length of the topological ES observed in other systems[13,14,17,27,28]. Furthermore, the conductance intensity of the ES exhibits the identical energy dependence and the equal lateral spatial distribution at the energies within the bulk-gap, as displayed in Fig. 2d, indicating the same origin of the 1D conducting channel along the whole edge of the $Bi_4I_4$ nanoribbon. When the bias voltage locates above the gap window, the 1D ES signal smears out and the 2D quasi-particle-interference (QPI) induced standing wave emerges in Fig. 2d and Supplementary Fig. 4. The wave-length of QPI patterns decreases with the increment of the bias energy, inferring electron-like band dispersion in this energy range. Therefore, we attribute the QPI patterns to the contribution of the conduction band and the peak of the black curve locating at 200 meV in Fig. 2b to the conduction band maximum of the 2D $Bi_4I_4$.

Compared to the trivial ES, the topological non-trivial ES in QSH insulators shows the robustness against to any of the non-TRS breaking perturbations. For trivial 1D ES, the point boundary at the end of the stripe is expected to evoke the standing wave similar to that of 2D QPI patterns due to the intervalley or intravalley scattering effects. Nevertheless, an ideal uniform distribution of 1D ES, especially along the straight stripe direction, is observed at all the energies within the gap, as shown in Fig. 2c and d, ruling out the LDOS intensity oscillations in conventional

1D electron system[29]. The scrutiny on the surface of the $Bi_4I_4$ nanoribbon in Fig. 2a implies the existence of adsorbates (marked by green dashed circle) and lattice distortion (marked by blue dashed circles). The STS mapping results in Fig. 2c and d indicate that the adsorbates have neglectable effect on the dispersion of our 1D ES, and the ES passes around the lattice distortion along the different corners without the obvious variation of penetration length and conductance intensity. It should be noted that the straight edges along the *b*-axis (stripe) direction are essentially distinct from other edges from the view of the existence of the dangling bond as the inter-stripe interaction is attributed to van der Waals force. Thus, the dangling bond effect on the 1D ES could be excluded. We also investigated the influence of "defect" on the 1D ES, as shown in Fig. 3. There are two kinds of line defects correlated to the absence of one stripe inside the $Bi_4I_4$ nanoribbon and close to the edge perpendicular to the stripe, which are marked by yellow circle and green circle in the topography (top image in Fig. 3), respectively. It is well known that the topological non-trivial ES is present in the boundary of topological insulators connecting with the trivial insulators. In our sample, the absent stripe could be treated as the trivial insulator - "vacuum" and performs as the medium of two QSH insulators. Therefore, the topological ES is anticipated to be presented surrounding the stripe-defects. In the spectroscopic mapping results, however, no ES is generated at the positions correlated to periphery of these two stripe-defects. Similar results have also been observed in QSH insulating layer with point defects in $Bi_{14}Rh_3I_9$[27], which may owe to large mean free path of the electrons in bismuth screening the medium vacuum[30]. All these experimental results are consistent with each other, and manifest the topological non-trivial origin of the 1D ES in $Bi_4I_4$ sample.

In order to confirm the topological non-trivial features and to illustrate the potential mechanism of the $Bi_4I_4$ on Bi(111) surface, we have studied their electronic properties by the first-principle calculations. Fig. 4a displays the schematic diagram of the 2D Brillion zones of free-standing monolayer $Bi_4I_4$ and $Bi_4I_4$ on Bi(111) surface, both of which have the same reciprocal vectors directions but are with different periodicities owe to the formation of superlattice in $Bi_4I_4$ on Bi(111). The calculated band structure of free-standing monolayer $Bi_4I_4$ without SOC, as shown in Supplementary Fig. 5, shows a direct band gap located at Γ point. From the view of density of states, both of the valence and conduction bands near the Fermi level are mainly contributed by Bi-6*p* orbitals, and the states dominated by I-4*p* orbitals are pushed far away from Fermi surface due to the larger electronegativity of I atoms. After the spin orbit coupling (SOC) is considered, the constituents and parities of the conduction and valence bands are inverted at Γ point, as shown in Fig. 4b. The

SOC-induced band inversion with the opposite parities indicates the non-trivial topological invariant in free-standing monolayer $Bi_4I_4$. Based on the evolutions of Wannier function centers in Brillion zone[31,32], we have calculated the $Z_2$ topological invariant within the first-principle framework, where the $Z_2 = 1$ verifies that the monolayer $Bi_4I_4$ belongs to the QSH insulator. More detailed information on the physical mechanism of the band inversion are listed in Supplementary Fig. 6. It is noticeable that the topological properties of monolayer $Bi_4I_4$ vary from topological semimetal to QSH insulator or normal insulator depending on various values of energy gap calculated by different methods[20,22], leading to the determination of the $Z_2$ invariants beyond the accuracy of theoretical calculations. Nevertheless, the solid experimental evidence for the topological properties of bulk $β$-$Bi_4I_4$ as a WTI has been revealed by laser-ARPES, which directly demonstrates the QSH insulator nature of monolayer $Bi_4I_4$. Therefore, our calculation results are reasonable and consistent with the experimental data in previous reports[22]. Since our $Bi_4I_4$ nanoribbons are synthesized on the Bi(111) surface, and the interface interaction modulate the crystalline structure of $Bi_4I_4$ to form the superlattice structure, the calculation of the electronic properties of $Bi_4I_4$ with the support of Bi(111) are performed. Fig. 4c displays the $Bi_{in}$-$p$ and $Bi_{ex}$-$p$ orbital projected character of bands with the effect of SOC strength, where the band and parities inversions at Γ point are retained. The theoretical simulations infer non-trivial topology of $Bi_4I_4$ on Bi(111) surface, in accordance with the measured experimental results.

Our study not only provides an experimental proof and the theoretical support to the topological ES in $Bi_4I_4$ nanoribbons, but also promotes the accessible scope of this 1D ES to the realization of the topological phase transition by the effects of strain and interlayer stacking order, which have been theoretically predicted or experimentally identified in previous reports[21,22,24,33]. This is of particular importance for situations that these factors are reasonably introduced into this system by using the nanoribbons as the building block. Furthermore, the long straight multiple conduction channels based on the quasi-1D nanoribbons are envisioned to be created for device applications through the technologically compatible lithographic patterning technique.

**Acknowledgements**

The work was supported by the Beijing Natural Science Foundation (Z180007), the National Natural Science Foundation of China (11874003, 11904015, 12004321, 12074021), the Natural Science Foundation of Hunan Province (No. 2019JJ50602), and National Key R&D Program of China (2018YFE0202700).

**Author contributions**

Y.N.L. and Y.D.L. carried out the experiments with the help of D.M. and Y.M. J.L. did the theoretical calculations. Y.N.L. and Y.D.L. grew the $Bi_4I_4$ nanoribbons. D.M., W.Z., and W.C.H. analyzed and interpreted the results. J.C.Z. supervised the project. J.C.Z., Y.D., and J.X.Z designed the experiment and wrote the manuscript.

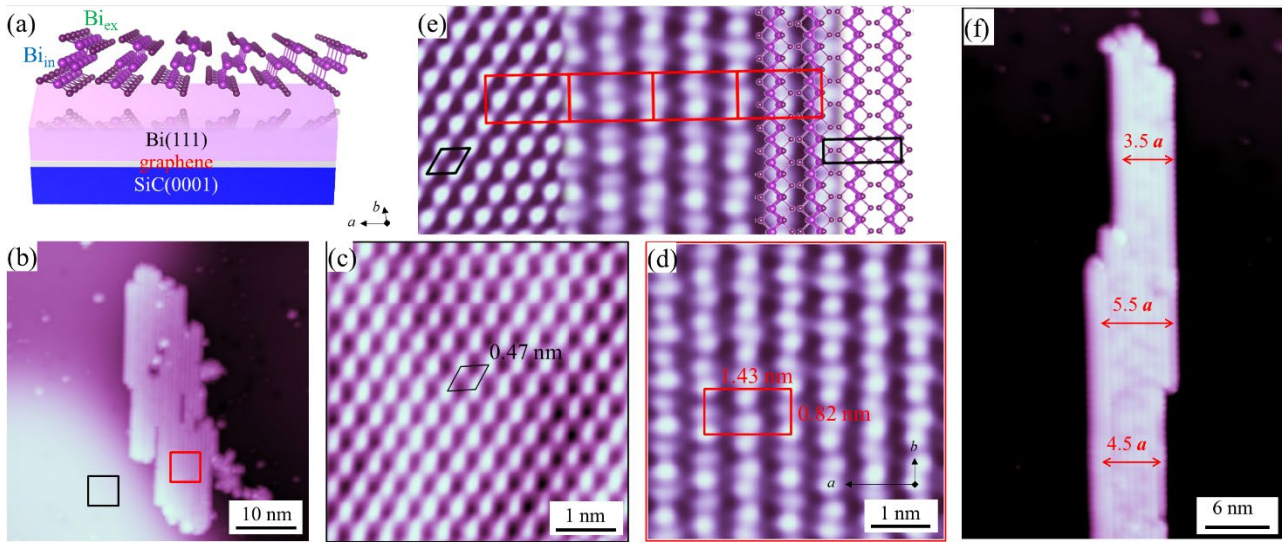

**Fig. 1** Bi$_4$I$_4$ nanoribbon with superlattice structure formed on the Bi(111) surface. **a** Illustration of Bi$_4$I$_4$/Bi(111)/graphene/SiC(0001) sandwich-like structure. Bismuth and iodine atoms are marked by purple and dark purple balls, respectively. Coordination polyhedra of two non-equivalent Bi atoms, Bi$_{in}$ (in the middle form a zigzag chain) and Bi$_{ex}$ atoms (bonded with Br at the edge of the chain) are indicated. **b** Large-sized STM topographic image of Bi$_4$I$_4$ nanoribbon on the surface of Bi(111) ($V_{bias}$ = 2.0 V, $I$ = 50 pA, 50 nm × 50 nm). **c**, **d** High-resolution STM images of **c** Bi(111) ($V_{bias}$ = 6 mV, $I$ = 4 nA, 5 nm × 5 nm) and **d** Bi$_4$I$_4$ ($V_{bias}$ = 5 mV, $I$ = 800 pA, 5 nm × 5 nm) measured in the area labeled by the black square and red square in **b**, respectively. The black rhombus and red rectangle are labelled for the unit cell of Bi(111) and 1 × 2 superlattice of Bi$_4$I$_4$, respectively. **e** Schematic diagrams of the evolution from the STM image of Bi(111) to STM image of Bi$_4$I$_4$ nanoribbon, and then to the atomic structural model of Bi$_4$I$_4$ (from left to right). The black rectangle represents the unit cell of 1 × 1 monolayer Bi$_4$I$_4$. **f** STM image of Bi$_4$I$_4$ nanoribbon with different width along the direction of $a$-axis ($V_{bias}$ = 2.0 V, $I$ = 50 pA, 30 nm × 50 nm). The "$a$" are marked for the lattice constant of Bi$_4$I$_4$ along $a$-axis.

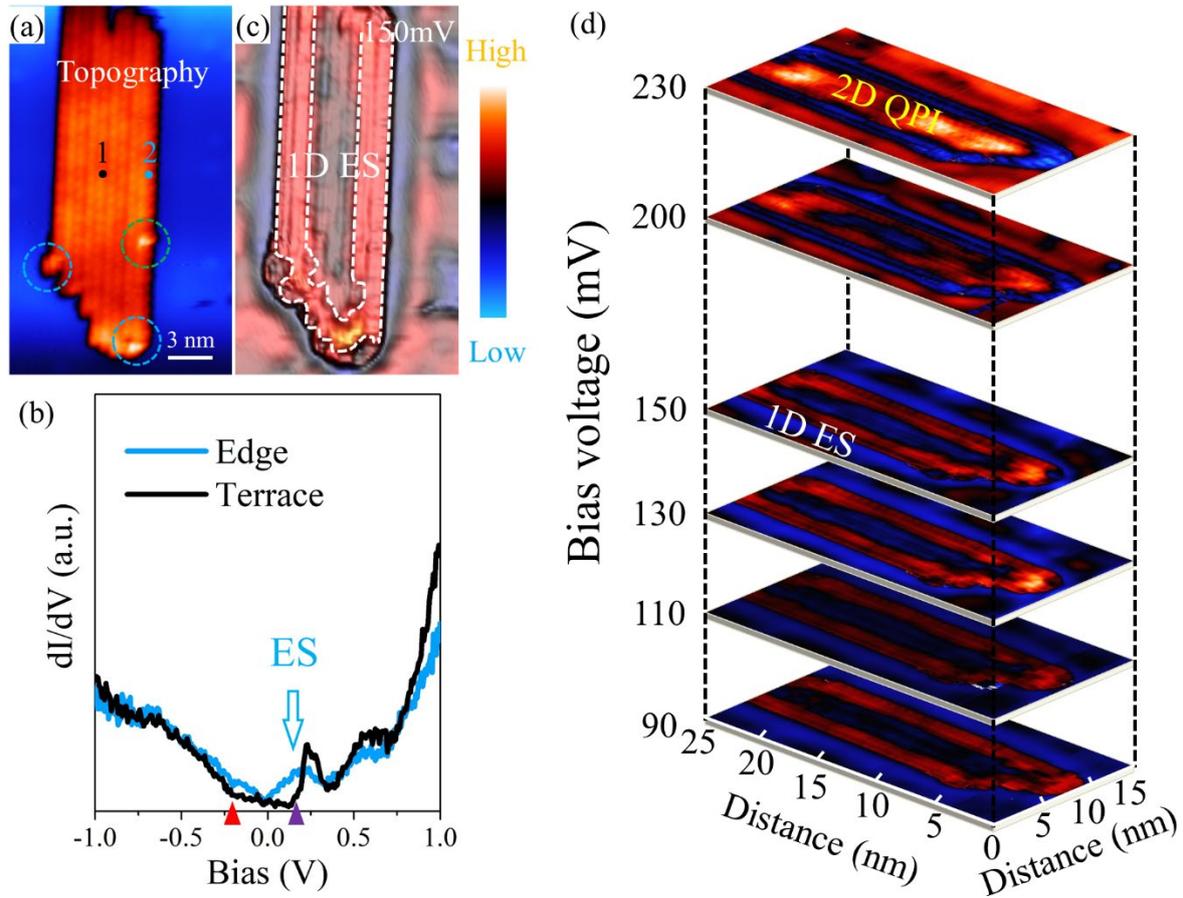

**Fig. 2** 1D edge states. **a** Topography of Bi$_4$I$_4$ nanoribbon ($V_{bias}$ = 0.15 V, $I$ = 300 pA, 15 nm × 25 nm). The blue dashed circles and green dashed circle represent the surface adsorbate and lattice distortion, respectively. **b** STS spectra obtained in the inner terrace (black) and edge (blue) regions of Bi$_4$I$_4$ nanoribbon in **a**. The red and purple triangles are presented for the valence band maximum (VBM) and conduction band minimum (CBM), respectively. **c** d$I$/d$V$ image collected with the bias at 150 mV of Bi$_4$I$_4$ nanoribbon in **a**. The white dashed lines are marked for the guidance of dispersion of 1D ES. **d** Stacked d$I$/d$V$ images of the area shown in **a** recorded at voltages across the bandgap as marked on the left.

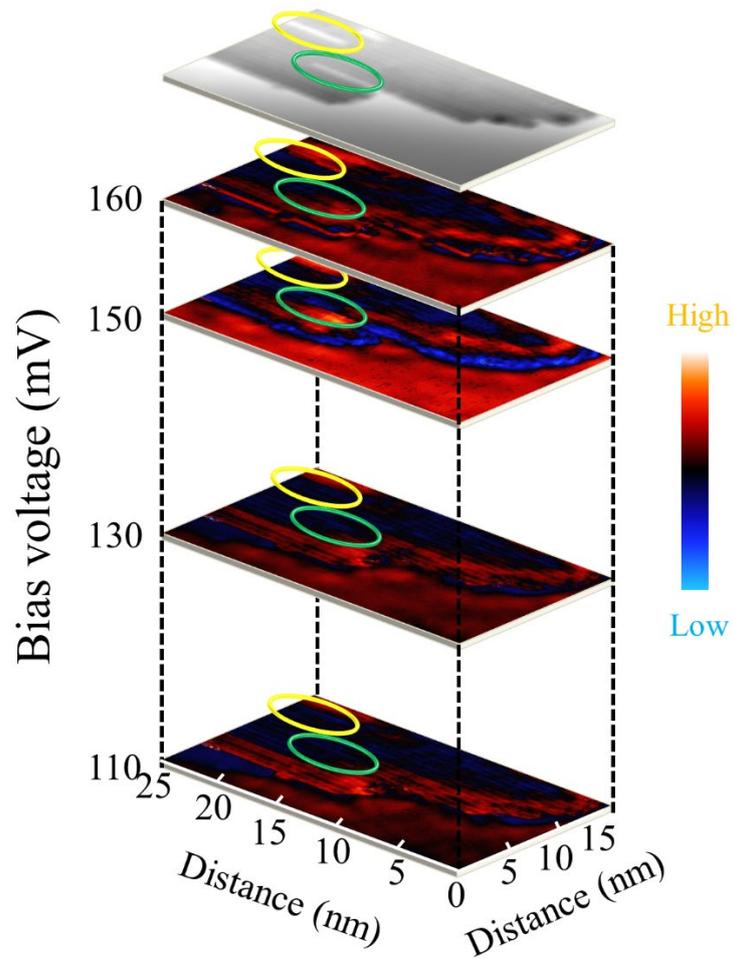

**Fig. 3** Non-trivial character of the ES: continuity in energy and space. Stacked d$I$/d$V$ images of the area of another Bi$_4$I$_4$ nanoribbon with the topography ($V_{bias}$ = 2.0 V, $I$ = 50 pA, 16 nm × 25 nm) shown in the top recorded at different voltages as marked on the left.

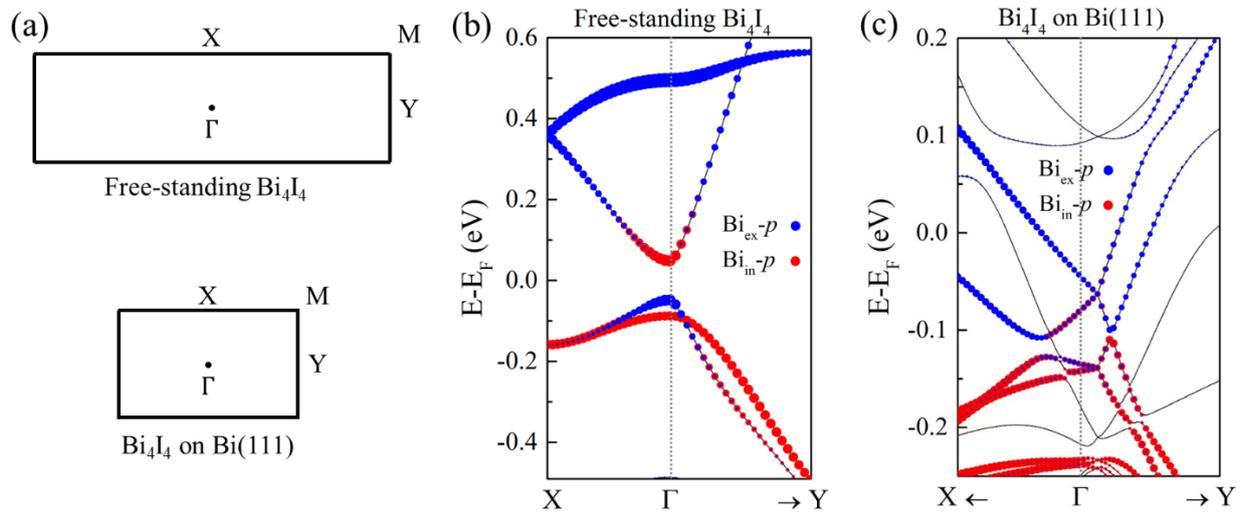

**Fig. 4** Calculated electronic properties with band and parities inversion in $Bi_4I_4$ system. **a** Schematic of 2D Brillouin zones of both free-standing monolayer $Bi_4I_4$ and on $Bi_4I_4$ Bi(111) surface with high symmetry points. **b**, **c** The $Bi_{in}$-$p$ and $Bi_{ex}$-$p$ orbital projected character of bands of **b** free-standing $Bi_4I_4$ and **c** $Bi_4I_4$ Bi(111) surface. The parity is labeled for CBM and VBM at the Γ-point.